\documentclass[fleqn,usenatbib]{mnras}
\usepackage{newtxtext,newtxmath}

\usepackage[T1]{fontenc}

\DeclareRobustCommand{\VAN}[3]{#2}
\let\VANthebibliography\thebibliography
\def\thebibliography{\DeclareRobustCommand{\VAN}[3]{##3}\VANthebibliography}

\usepackage{graphicx}	
\usepackage{amsmath}	
\usepackage{caption}
\usepackage{subcaption}
\usepackage{comment}
\usepackage{placeins}

\newcommand{\au}{\, {\rm au}}
\newcommand{\me}{\, {\rm M}_{\oplus}}
\newcommand{\Stokes}{{\, {\rm St}}}

\title[External photoevaporation and pebble accretion ]{Planet formation via pebble accretion in externally photoevaporating discs}

\author[L. Qiao et al.]{
Lin Qiao,$^{1}$\thanks{E-mail: lin.qiao@qmul.ac.uk}
Gavin A. L. Coleman,$^{1}$
Thomas J. Haworth$^{1}$
\\
$^{1}$Astronomy Unit, School of Physics and Astronomy, Queen Mary University of London, London E1 4NS, UK\\
}

\date{Accepted XXX. Received YYY; in original form ZZZ}

\pubyear{2021}

\begin{document}
\label{firstpage}

\pagerange{\pageref{firstpage}--\pageref{lastpage}}
\maketitle

\begin{abstract}
We demonstrate that planet formation via pebble accretion is sensitive to external photoevaporation of the outer disc. In pebble accretion, planets grow by accreting from a flux of solids (pebbles) that radially drift inwards from the pebble production front. If external photoevaporation truncates the outer disc fast enough, it can shorten the time before the pebble production front reaches the disc outer edge, cutting off the supply of pebble flux for accretion, hence limiting the pebble mass reservoir for planet growth. Conversely, cloud shielding can protect the disc from strong external photoevaporation and preserve the pebble reservoir. Because grain growth and drift can occur quickly, shielding even on a short time-scale ($<1$\,Myr) can have a non-linear impact on the properties of planets growing by pebble accretion. For example a $10^{-3}\,M_\oplus$ planetary seed at 25\,au stays at 25\,au with a lunar mass if the disc is immediately irradiated by a $10^3$\,G$_0$ field, but grows and migrates to be approximately Earth-like in both mass and orbital radius if the disc is shielded for just 1\,Myr. In NGC 2024, external photoevaporation is thought to happen to discs that are $<0.5$\,Myr old, which coupled with the results here suggests that the exact planetary parameters can be very sensitive to the star forming environment. Universal shielding for time-scales of at least $\sim1.5\,$Myr would be required to completely nullify the environmental impact on planetary architectures. 
\end{abstract}




\section{Introduction}
There are now more than 5000 confirmed exoplanets exhibiting very diverse properties and planetary system architectures \citep[e.g.][]{2021exbi.book....2G}. Understanding this diversity of planets requires understanding planet formation and evolution. However this is not straightforward as planet formation involves the complex interplay of many processes \citep[see reviews by e.g.][]{2022arXiv220309759D, 2022ASSL..466....3R,2023arXiv230300012E}. One promising model for the formation of terrestrial planets and cores of giant planets (in the core accretion scenario) is pebble accretion, in which mm--cm sized pebbles are mainly responsible for planetary growth by their accretion onto larger planetesimals and cores \citep{2012A&A...544A..32L,2014A&A...572A.107L}. The pebble accretion mechanism is an efficient way of resolving issues associated with the traditional core accretion model: the growth barrier when particles reaches mm-cm due to radial drift \citep{1977MNRAS.180...57W}; and slow planetesimal accretion with multi-kilometer sized bodies \citep{Pollack,2014MNRAS.445..479C,ColemanNelson16,ColemanNelson16b}. Pebbles are accreted on to growing planetary embryos at a much faster rate than planetesimals due to the greater kinetic energy dissipation by gas drag when pebbles enter the gravitational reach of a core, \citep{2010A&A...520A..43O, 2012A&A...544A..32L}. There has also been some success in being able to form compact terrestrial planetary systems such as Trappist-1 via pebble accretion \citep{2017A&A...604A...1O, Coleman19}.

One potential factor contributing to planet diversity is the birth environment. Most stars form in large stellar clusters with high stellar density \citep[]{2003ARA&A..41...57L, 2008ApJ...675.1361F, 2020MNRAS.495L..86L, 2020MNRAS.491..903W}, and there has been growing attention on how the dense cluster environment can influence the planet-forming disc evolution and possibly the final planetary systems. One way the stellar cluster environment can influence a protoplanetary disc is via star-disc gravitational encounters \citep[for a recent review see][]{2023EPJP..138...11C}. The other environmental impact on discs is due to ``external'' UV irradiation by massive stars in the cluster \citep[for a recent review see][]{2022EPJP..137.1132W}. This external irradiation drives  a thermal wind from the outer disc in a process called external photoevaporation. In addition to depleting the disc mass, if material from the outer disc is removed by external photoevaporation more quickly than its resupply via viscous spreading, then the disc can be truncated \citep{2007MNRAS.376.1350C, 2017MNRAS.468.1631R, 2018ApJ...860...77E,2022MNRAS.514.2315C}. 

There have been numerous direct observations of discs being affected by their environments. When the radiation field is sufficiently strong, the circumstellar disc is enshrouded in a cometary wind, with the cusp directed towards the exciting UV source. These ``proplyds'' have been observed across a range of UV radiation field environments in the Orion Nebular Cluster and NGC 1977 \citep[e.g.][]{1994ApJ...436..194O,1999AJ....118.2350H, 2012ApJ...756..137B, 2016ApJ...826L..15K}. There is also recent evidence of external photoevaporation happening at very early times in NGC 2024 \citep[$<0.5$\,Myr ,][]{2021MNRAS.501.3502H}, competing with even the earliest evidence for planet formation \citep{2018ApJ...857...18S, 2020Natur.586..228S}. One disc in $\sigma$ Ori is observed with unusually high [OI] 6300\AA\ emission, which has also been interpreted as due to an external photoevaporative wind \citep{1998ApJ...502L..71S, 2009A&A...495L..13R,2023MNRAS.518.5563B}. Additionally, statistical comparisons of disc properties in clusters such as masses \citep{2017AJ....153..240A, 2018ApJ...860...77E}, radii \citep{2018ApJ...860...77E, 2020ApJ...894...74B, 2021ApJ...923..221O} and disc fractions \citep{2019MNRAS.486.4354R, 2020A&A...640A..27V} also show evidence of external photoevaporation. 

Analytic and semi-analytic models of the external photoevaporation of discs were developed after the discovery of proplyds \citep{1998ApJ...499..758J, 1999ApJ...515..669S, 2004ApJ...611..360A}. \citet{2007MNRAS.376.1350C} coupled viscous evolution of discs with mass loss rate calculations from semi-analytical models of photoevaporation. However, it is specifically far-ultraviolet (FUV) light that typically drives external photoevaporative winds, so to accurately calculate the mass loss rate it is necessary to determine the temperature in the photodissociation region (PDR) from microphysics which cannot be solved analytically. \citet{2018MNRAS.475.5460H} developed a publicly available grid of mass loss rates (called FRIED) computed with PDR-hydrodynamics simulations, which has enabled the community to run on-the-fly modelling of disc evolution with external photoevaporation \citep[e.g.][]{2018MNRAS.481..452H, 2019MNRAS.485.1489W, 2019MNRAS.490.5478W, 2020MNRAS.497L..40W, 2020MNRAS.491..903W,2019MNRAS.490.5678C, 2020MNRAS.492.1279S, 2021MNRAS.501.1782C, 2022MNRAS.tmp.1811C, 2021MNRAS.502.2665P, 2022MNRAS.512.3788Q,  2023arXiv230203721W}. Studies that included both effects of dynamical encounters and external photoevaporation of discs in dynamically evolving stellar clusters \citep{2001MNRAS.325..449S, 2018MNRAS.478.2700W, 2019MNRAS.490.5678C, 2021MNRAS.501.1782C, 2022MNRAS.tmp.1811C, 2023arXiv230203721W} generally conclude that external photoevaporation is overall more significant in terms of causing mass loss, but both mechanisms are likely to impact disc evolution, and the interplay between these processes is still not well understood. In this paper we focus on planet formation in the context of external photoevaporation, but the additional role of encounters should be included in future.



In recent years, there have been efforts to more directly couple disc evolution calculations to star formation simulations. For example, \citet{2019MNRAS.485.1489W}, \cite{2019MNRAS.490.5678C}, \cite{2021MNRAS.501.1782C}, \cite{2021ApJ...913...95P} and \cite{2021ApJ...913...95P} coupled n-body simulations of stellar cluster dynamics with external photoevaporation and disc evolution models. In those calculations all stars/discs are assumed to have formed at the start of the simulation. They calculate the time varying FUV field incident upon each disc (and hence the time varying external photoevaporative mass loss rate from FRIED) due to the aggregate of stars in the cluster, assuming geometric dilution of each star's intrinsic FUV luminosity. However this does not account for  ongoing star formation, shielding by the star forming cloud or feedback processes such as ionising radiation, winds and supernovae that can act to disperse the cloud. \citet{2022MNRAS.tmp.1811C} coupled external photoevaporation with a simulation following the collapse of a giant molecular cloud in an SPH hydrodynamical model that self-consistently formed stars, and found that ongoing star formation is necessary for massive discs to exist beyond the early cluster age. Recently \citet{2022MNRAS.512.3788Q} and \citet{2023arXiv230203721W}
coupled star cluster formation and feedback simulations with disc evolution models. These directly account for the impact of cloud shielding on the UV field incident upon discs, and the dispersal of the shielding matter over time by feedback. They found that the effect of shielding in dense stellar clusters can prevent high external photoevaporative mass loss at least in the early stage of disc evolution. However, typical shielding time-scales will depend on the nature of the star formation event, such as the cloud virial ratio, which was already known to determine how effectively and rapidly feedback can disperse the cloud \citep[][]{2012MNRAS.424..377D, 2012MNRAS.427.2852D, 2013MNRAS.430..234D, 2021MNRAS.502.2665P,  2022MNRAS.512.3788Q,  2023arXiv230203721W}. 

Overall, it is now well recognised that in a evolving star forming region, external photoevaporation and cloud shielding times can have significant impacts on disc evolution. However, what is much less studied is the impact of external photoevaporation on planet formation. Recently, \cite{2022MNRAS.515.4287W} found that the gas accretion and migration of wide-orbit giant planets in a disc can be suppressed by FUV-induced photoevaporation. There are also population synthesis models incorporating mass loss in gas (interpolated from FRIED grid) and dust entrained through external photoevaporative winds  \citep[e.g.][]{2022A&A...666A..73B, 2023arXiv230300012E}.  \citet{2022A&A...666A..73B} found that dust entrainment in photoevaporative winds can be an important mechanism for removing solids. However, one unclear but important aspect of all this is the competition between early, planet formation \citep[substructures in 0.5\,Myr discs suggests possible early planet formation][]{2018ApJ...857...18S, 2020Natur.586..228S} and external photoevaporation. If a planet grows via pebble accretion, external photoevaporation could possibly limit the inwardly drifting pebble mass budget by depleting the outer disc. Conversely, cloud shielding could protect the disc for some time before being exposed to a high radiation field \citep{2022MNRAS.512.3788Q,  2023arXiv230203721W} which could give planet formation a head start. Therefore the impact of external photoevaporation and shielding might affect not just the formation of wide-orbit giant planets, but also compact planetary systems. We therefore need to understand how important the shielding time-scale is for planet formation theoretically, and to observationally determine how long it is in practice. The objective of this paper is to provide the theoretical component. We couple a model of planet formation via pebble accretion in a viscously evolving disc with external photoevaporation induced by a parameterised time-varying FUV radiation field. We aim to isolate the impacts of of external photoevaporation and shielding on planet formation by pebble accretion.

\section{Numerical Method}

We investigate the formation of planets via pebble accretion in externally photoevaporating protoplanetary discs.
To isolate the impact of external photoevaporation and cloud shielding on the planet formation process in a controlled way we apply a state of the art planet formation code with parameterised time-varying external irradiation field. The disc evolves viscously (section \ref{subsect_discevol}) and is also subject to mass loss via external photoevaporation due to the parameterised time varying external FUV radiation field strength (section \ref{FUV_track_section}). For simplicity, in each calculation we only consider a single planetary embryo in the disc. This can then evolve by accreting pebbles (sect. \ref{sub_sec_pebacc}) and a gaseous envelope (sect. \ref{sub_sect_gasacc}) while migrating in the type I migration regime (sect. \ref{subsect_typeImig}).

In total, we vary three parameters in this study: initial shielding time $t_{\textrm{sh}}$ (0 - 3 Myrs), the maximum FUV field strength that discs are exposed to after the initial shielding time $F_{\textrm{FUV, max}}$ (10 - 10000 $G_{0}$) and the initial semi-major axis of planetary embryos $a_{\textrm{init}}$ (10 - 100 AU). Each simulation has just a single planetary embryo in the disc, and is run for a total of 10 Myrs, unless the planet migrates interior to 1$\au$ or the disc mass drops below 0.1\,M$_\oplus$. For the purpose of isolating the impact of external photoevaporation, we do not consider internal photoevaporation at this stage (see section \ref{external_evap} for detailed discussions.)

\subsection{Disc Evolution}
\label{subsect_discevol}
We adopt a 1D viscous disc model where we evolve the gas surface density by solving the standard diffusion equation with an additional term to describe mass loss by external photoevaporation:
\begin{equation}
\dot{\Sigma}(r) = \frac{1}{r}\frac{d}{dr}\left[ 3 r^{1/2}\frac{d}{dr}(\nu \Sigma r^{1/2}) \right] - \dot{\Sigma}_{PE}(r),
\label{eqn:Surface_dens}
\end{equation}
where $\dot{\Sigma}_{PE}(r)$ is the rate of change of surface density due to external photoevaporative wind induced by FUV radiation (see section \ref{external_evap}). We use the standard $\alpha$ model for the disc viscosity \citep{1973A&A....24..337S}
\begin{equation}
    \nu = \alpha {c_{s}}^{2} / \Omega, 
\end{equation}
where $c_s$ is the local sound speed, $\Omega$ is the angular velocity and $\alpha$ is the viscosity parameter, which is set at $\alpha = 10^{-3}$ for all simulations. 

For the initial gas surface density profile, we use the similarity solution of \citet{1974MNRAS.168..603L}

\begin{equation}
\label{sigma_initial}
    \Sigma = \Sigma_{0}\left(\frac{R}{R_C}\right)^{-1}\exp{\left(-\frac{R}{R_C}\right)},
\end{equation}
where $\Sigma_0$ is the normalisation constant set by the total disc mass (for a given $R_C$), and $R_C$ is the scale radius, which sets the initial disc size. For all our simulations, we use an initial disc mass of 105 $M_{\textrm{Jup}}$ ($\sim 0.1 M_{*}$) and an initial scale radius $R_C$ = 50 AU. We also use a constant host star mass $M_{*}$ = 1 M$_\odot$\ and solve equation \ref{eqn:Surface_dens} over 2000 grid cells using an explicit finite difference scheme, adopting a non-uniform mesh grid, for which the grid spacing $\Delta r$ scales with radius \citep{2014MNRAS.445..479C}. We set the local solids-to-gas ratio ($Z_0$) everywhere in the disc to 0.01 and note that the solids in the disc includes both dust particles that contribute to the disc opacity and pebbles produced from dust coagulation (see section \ref{sub_sec_pebacc}).

We set an initial radial temperature profile as: 
\begin{equation}
    T_\textrm{{initial}} = T_{0}\left(\frac{R}{R_0}\right)^{-0.5}, 
\end{equation}
where $R_0 = 1$\,au and $T_{0}$ is the temperature at $R_0$ (1\,au), which is set to $280K$. The disc temperature after each time step of the simulation is obtained by solving the following thermal equilibrium equation which describes the balance of heating from central star irradiation, heating from residual molecular cloud, viscous heating, and blackbody cooling using an iterative method:
\begin{equation}
    Q_\textrm{{irr}} + Q_{\nu} + Q_\textrm{{cloud}} - Q_\textrm{{cool}} = 0,
\end{equation}
where $Q_{\textrm{irr}}$ is the radiative heating rate due to the central star, $Q_{\nu}$ is the viscous heating rate per unit area of the disc, $Q_{\textrm{cloud}}$ is the radiative heating due to the residual molecular cloud (with temperature 10K), and $Q_{\textrm{cool}}$ is the blackbody radiative cooling rate \citep[see equations 4-10 in][for further descriptions of the expressions in calculating the temperature]{2021MNRAS.506.3596C}. For reference, in the Appendix Figure \ref{fig:surface_density} we show the surface density evolution for discs in different UV field strengths and with different shielding times.

\subsubsection{External photoevaporation}
\label{external_evap}
For a disc in a dense stellar cluster, external photoevaporative winds can be launched by both EUV and FUV photons radiated from nearby massive stellar neighbours, though the FUV is generally expected to be dominant in setting the mass loss rate \citep{1994ApJ...436..194O, 1998ApJ...499..758J, 2004ApJ...611..360A, 2016MNRAS.457.3593F, 2022EPJP..137.1132W}. 

In our simulations, the mass loss rate due to external photoevaporation is calculated via interpolating over the FRIED grid \citep{2018MNRAS.481..452H}. FRIED grid provides mass loss rates for discs irradiated by FUV radiation as a function of the star/disc/FUV parameters. In our simulations, we determine the mass loss rate at each time step by linearly interpolating FRIED  in three dimensions: disc size $R_{d}$, outer edge disc surface density $\Sigma_{\textrm{out}}$ and FUV field strength at this time $F_{\textrm{FUV}}(t)$ (see section \ref{FUV_track_section} for details about the FUV field tracks used.)



A cautionary note is that  FRIED mass loss rates are sensitive to the value of the disc outer radius $R_{d}$ and the surface density there $\Sigma_{\textrm{out}}$. However,  towards the disc outer edge the surface density can be a strong function of radius (for example dropping off exponentially). Furthermore, if the disc outer radius is chosen at a point that would be optically thin to the external irradiation, the situation is unphysical since in reality radiation would propagate deeper into the disc and drive the wind from there. This means the disc outer edge location should be at the optically thick/thin transition, where the interpolated mass loss rate from FRIED is a maximum. We therefore adopt that radius of maximum mass loss $R_{\textrm{max}}$ as the disc outer radius \citep[see ][particularly figure 2 for more information]{2020MNRAS.492.1279S}.

We evaluate the FRIED mass loss rate at each radius in the disc to determine $R_{\textrm{max}}$. We then extract the corresponding mass loss from the disc via

\begin{equation}
    \dot{\Sigma}_{\textrm{ext, FUV}}(r) = G_{\rm sm} \frac{\dot{M}_\textrm{{ext}}(R_{\textrm{\textrm{max}}})}{\pi(R^2_\textrm{{d}} - {R_{\textrm{\textrm{max}}}}^2)+A_{\rm sm}}, 
\end{equation}
where $A_{\rm sm}$ is a smoothing area equal to 
\begin{equation}
A_{\rm sm} = \dfrac{\pi(R_{\rm max}^{22}-(0.1 R_{\rm max})^{22})}{11R_{\rm max}^{20}}
\end{equation}
and $G_{\rm sm}$ is a smoothing function
\begin{equation}
    G_{\rm sm} = \dfrac{r}{R_{\rm max}^{20}}.
\end{equation}

Note that at this stage, to isolate the impact of external photoevaporation, we do not include internal photoevaporative winds \citep[e.g.][for reviews]{2017RSOS....470114E, 2022EPJP..137.1357E}. Internal photoevaporative winds can entrain small dust in the wind \citep{2021MNRAS.501.1127H,2016MNRAS.461..742H,2016MNRAS.463.2725H} and eventually open up an inner hole on a time-scale that depends on the inwards transport of gas and the strength of the radiation field (EUV, FUV, or X-ray) driving the wind \citep[e.g.][]{2022MNRAS.514..535S,2010MNRAS.401.1415O, 2012MNRAS.422.1880O, 2019MNRAS.487..691P,2021MNRAS.508.3611P,2010MNRAS.401.1415O,2022MNRAS.514.2315C}. Studying the combined interplay with these processes is left for future work. 

\subsubsection{Shielding time-scale prescription}
\label{FUV_track_section}
\begin{figure}
    \centering	\includegraphics[width=\columnwidth]{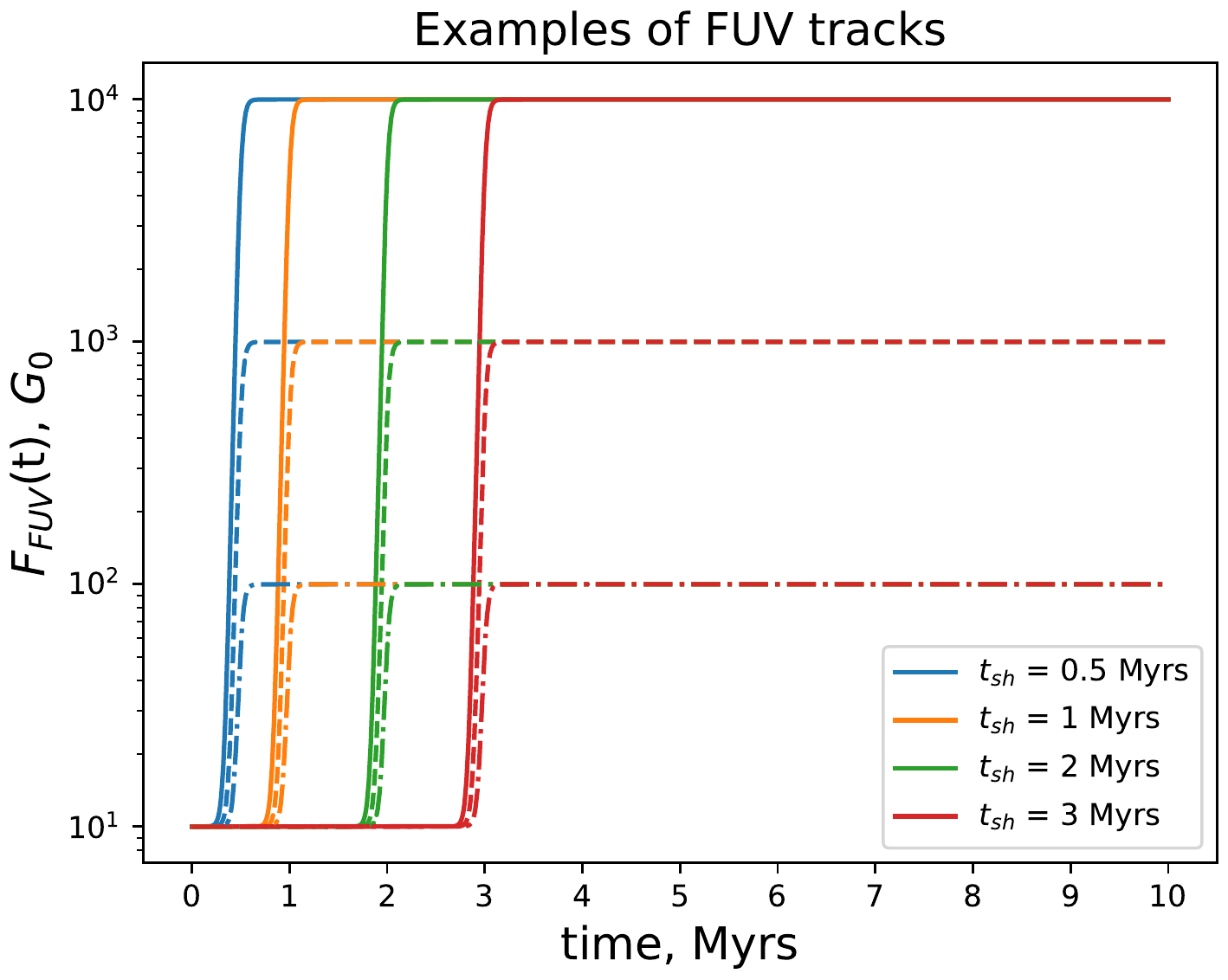}
    \caption{Examples of some of the time varying FUV tracks used in the simulation described by equation \ref{eqn:FUV_track}, the colours indicate different shielding times, with blue, orange, green and red representing $t_{\textrm{sh}} = $ 0.5, 1, 2, 3 Myrs, and the line-styles represent different values of ${F_{\textrm{FUV, max}}}$, the solid curves represent $F_{\textrm{FUV, max}} = 10000$ $G_0$, the dashed curves represent $F_{\textrm{FUV, max}} = 1000$ $G_0$, dash-dotted lines represent $F_{\textrm{FUV, max}} = 100$ $G_0$.} 
    \label{fig:FUV_tracks}
\end{figure}

We aim to investigate the effect of shielding time $t_{\textrm{sh}}$ on planet formation by pebble accretion in a controlled way. Therefore, rather than follow \cite{2022MNRAS.512.3788Q} or \cite{2023arXiv230203721W} and use complicated time varying FUV radiation fields from simulations of star formation, here we opt to use a parameterised time varying FUV flux profile that transitions from a low FUV field $F_{\textrm{FUV, 0}}$ to a high value $F_{\textrm{FUV, max}}$ after a shielding period $t_{\textrm{sh}}$
\begin{multline}
\label{eqn:FUV_track}
    F_{\textrm{FUV}}(t) = F_{\textrm{FUV,0}} + \frac{1}{2}(F_{\textrm{FUV,max}}  - F_{\textrm{FUV,0}}) \\ 
    \times \left(\tanh \left(\frac{t - t_\textrm{{shield}}}{t_{\textrm{trans}}}\right) + 1\right),
\end{multline}
where $t_{\textrm{trans}}$ = $5\times10^4$\,years is a parameter that controls the time-scale over which the star/disc moves from a shielded to an unshielded environment. This rapid transition from shielded to irradiated represents the change in irradiation when stars/discs cease being embedded \citep{2022MNRAS.512.3788Q}. More complicated time variation of the UV could arise due to cluster dynamics (e.g. a star moving from a low UV region into a higher UV region), but including this in the present study makes the parameter space unwieldy. We set $F_{\textrm{FUV, 0}}$ = 10 $G_{0}$, and vary $t_{\textrm{sh}}$ (0 - 3 Myrs) and $F_{\textrm{FUV, max}}$ (10 - 10000 $G_{0}$) as parameters in our simulations. Figure \ref{fig:FUV_tracks} illustrates some examples of our time varying external FUV radiation parameterisation. 

\subsection{Pebble accretion}
\label{sub_sec_pebacc}

We use the model of pebble production and  accretion implemented in \citet{2021MNRAS.506.3596C}, which follows the models of \citet{2012A&A...544A..32L} and \citet{2014A&A...572A.107L}. In this model of pebble production via dust growth, solids are separated into two populations: one of radially stationary small dust grains and one of larger inwardly drifting pebbles, formed from the coagulation of the vertically settling small dust grain population in the disc. At each disc radius it takes a certain amount of time for the small ISM-like dust grains \citep[about 0.1 - 1 $\mu$m ][]{2001ApJ...548..296W} to grow to a pebble size large enough to start drifting $R_{\textrm{drift}}$ ($R_{\textrm{drift}}$ $\approx$ 1 - 10mm \citep{2014A&A...572A.107L}), assuming a certain growth efficiency. In other words at a certain time $t$, there is one particular disc radius $r_{g}(t)$ which is called the pebble production front, where the dust particles have just grown to $R_{\textrm{drift}}$:
\begin{equation}
\label{eqn:pebble_front}
    r_{g}(t) = {\left(\frac{3}{16}\right)}^{1/3}{(G M_{*})}^{1/3}{(\epsilon_{d} Z_{0})}^{2/3}t^{2/3},
\end{equation}
where $\epsilon_{d} = 0.05$ defines the size-dependent dust growth efficiency, and $Z_0 = 0.01$ is the solids-to-gas ratio, contributed by both pebbles ($Z_{\textrm{peb}}$) and dust ($Z_{\textrm{dust}}$):
\begin{equation}
    \label{eqn:Z}
    Z_{0} = Z_{\textrm{peb}} + Z_{\textrm{dust}}.
\end{equation}
In this work we assume that 90\% of the total solids is converted into pebbles, and this conversion rate remains constant throughout the simulation. 

As the dust growth time-scale is shorter at smaller disc radii (i.e. dust at smaller radii grows faster), the pebble production front moves outwards from the disc inner edge over time (see figure \ref{fig:pebbles_cutofftime} for the plot of $r_{g}(t)$ for the disc configuration used in this paper).
This provides a pebble mass flux (flux of pebbles drifting inwards from $r_{g}$) defined as:
\begin{equation}
\label{eqn:pebble_flux}
    \dot{M}_{\textrm{flux}} = 2 \pi r_g \frac{dr_{g}}{dt} Z_{\textrm{peb}}(r_g) \Sigma_{\textrm{gas}}(r_g),
\end{equation}
where $\Sigma_{\textrm{gas}}(r_g)$ is the gas surface density at the pebble production front. From the mass flux we define the pebble surface density profile $\Sigma_{\textrm{peb}}$ the same way as \citet{2014A&A...572A.107L}:
\begin{equation}
    \Sigma_{\textrm{peb}} = \frac{\dot{M}_{\textrm{flux}}}{2 \pi r v_{r}},
\end{equation}
where $r$ is the disc radius and $v_r$ is the radial velocity of pebbles at $r$, defined as:
\begin{equation}
    v_r = 2 \frac{\Stokes}{\Stokes^{2} + 1} \eta v_{k} - \frac{v_{\textrm{r,gas}}}{1+ \Stokes^{2}}
\end{equation}
\citep{1977MNRAS.180...57W,1986Icar...67..375N} ,where $St$ is the Stokes number of the pebbles, $v_k$ is the local keplerian velocity, $v_{\textrm{r,gas}}$ is the local gas radial velocity, and $\eta$ is the dimensionless measure of gas pressure support:
\begin{equation}
\eta = -\frac{1}{2}{\left(\frac{H}{r}\right)}^{2}\frac{\partial \textrm{ln}P}{\partial \textrm{ln}r},
\end{equation}
where $H$ is the disc scale height \citep{1986Icar...67..375N}.
For the Stokes number, we assume it is equal to:
\begin{equation}
    \Stokes = \min(\Stokes_{\rm drift}, \Stokes_{\rm frag})
\end{equation}
where $\Stokes_{\rm drift}$ is the drift-limited Stokes number that is obtained through an equilibrium between the drift and growth of pebbles to fit constraints of observations of pebbles in protoplanetary discs and from advanced coagulation models \citep{2012A&A...539A.148B}
\begin{equation}
    \Stokes_{\rm drift} = \dfrac{\sqrt{3}}{8}\dfrac{\epsilon_{\rm p}}{\eta}\dfrac{\Sigma_{\rm peb}}{\Sigma_{\rm gas}},
\end{equation}
where $\epsilon_{\rm p}$ is the coagulation efficiency between pebbles which we assume is similar to the dust growth efficiency and is equal to 0.5.
As well as the drift-limited Stokes number, we also include the fragmentation-limited Stokes number, ($\Stokes_{\rm frag}$) which we follow \citet{Ormel07} and is equal to
\begin{equation}
    \Stokes_{\rm frag} = \dfrac{v_{\rm frag}^2}{3\alpha c_{\rm s}^2}
\end{equation}
where $v_{\rm frag}$ is the impact velocity required for fragmentation, which we model as the smoothed function
\begin{equation}
    \dfrac{v_{\rm frag}}{1 \rm ms^{-1}} = 10^{0.5+0.5\tanh{((r-r_{\rm snow})/5H)}},
\end{equation}
where $r_{\rm snow}$ is the water snowline, which we assume to be where the disc midplane temperature is equal to 170K. The fragmentation velocity therefore varies between $1$\,ms$^{-1}$ for rocky pebbles \citep{Guttler10}, to $10$\,ms$^{-1}$ for icy pebbles, consistent with some results in the literature \citep[though this is still an area of open research][]{Gundlach15,Musiolik19}.

For each simulation, we inject the planetary embryo core at its specified initial semi-major axis $a_{\textrm{init}}$ as soon as $r_{g}$ has reached $a_{\textrm{init}}$ (i.e. we inject it at $t$ when $r_{g} = a_{\textrm{init}}$) with the initial planetary embryo core mass $M_{\textrm{core, init}}$ defined as 10\% of the transition mass $M_{\textrm{trans}}$ for the core. The transition mass is a core mass threshold above which the pebble accretion onto the core transitions from the Bondi regime to the Hill regime \citep{2017AREPS..45..359J}. The Bondi accretion regime happens initially for low embryo core mass, when the core accretes pebbles that pass through its Bondi radius $R_B$, which is smaller than its Hill radius $R_H$ at low mass. As the core becomes more massive and reaches $M_{\textrm{trans}}$, its Bondi radius becomes similar to the Hill radius, and the accretion regime switches to the Hill regime where the accretion rate is limited by its Hill sphere. The transition mass is calculated as:
\begin{equation}
M_{\textrm{trans}} = \eta^{3} M_{*}.
\end{equation}
\citet{2021MNRAS.506.3596C} studied the sizes and distributions of planetary embryos and planetesimals formed in discs via pebble trapping in short-lived pressure bumps, and found that the masses of the planetary embryos formed are at least one order of magnitude lower than the transition mass, which is the reason we chose our initial embryo core masses as 0.1 $M_{\textrm{trans}}$. 

As in \citet{2021MNRAS.506.3596C}, we use two modes for core growth via pebble accretion: a 2D or a 3D mode, depending on the core's Hill radius $R_H$ and the scale height of the pebbles $H_{\textrm{peb}}$. If $R_{\textrm{H}} < H_{\textrm{peb}}$, we use the 3D accretion mode since there are pebbles crossing the core's entire Hill sphere, but as the core mass increases and when $R_{\textrm{H}} > H_{\textrm{peb}}$, we switch to the 2D accretion mode as some regions of its Hill sphere are empty of pebbles. We use the 2D and 3D accretion rates in \citet{2017AREPS..45..359J}:
\begin{equation}
    \dot{M}_{\textrm{2D}} = 2 R_{\textrm{acc}} \Sigma_{\textrm{peb}} \delta v,
\end{equation}
and 
\begin{equation}
    \dot{M}_{\textrm{3D}} = \pi {R_{\textrm{acc}}}^{2} \rho_{\textrm{peb}} \delta v,
\end{equation}
where $\rho_{\textrm{peb}}$ is the midplane pebble density, $\delta v=\Delta v + \Omega R_{\textrm{acc}}$ is the approach speed, with $\Delta v$ as the gas sub-Keplerian velocity. $R_{\textrm{acc}}$ is the accretion radius which depends on whether the accreting core is in the Hill or Bondi regime, and on the pebble friction time-scale. The dependency on friction time arises due to pebbles having to change directions significantly on time-scales shorter than the friction time-scale, which brings a criterion accretion radius $\hat{R}_{\textrm{acc}}$ defined as:
\begin{equation}
    \hat{R}_{\textrm{acc}} = {\left( \frac{4t_f}{t_B} \right)}^{1/2} R_B
\end{equation}
for the Bondi regime, where $t_f = \Stokes/\Omega $ is the pebble friction time-scale, $t_B$ is the Bondi sphere crossing time, and
\begin{equation}
    \hat{R}_{\textrm{acc}} = {\left( \frac{\Omega_{k} t_f}{0.1} \right)}^{1/2} R_H
\end{equation}
for the Hill regime. The accretion radius is the defined as:
\begin{equation}
    R_{\textrm{acc}} = \hat{R}_{\rm acc} \exp{ \left [ - \chi {(t_f/t_p)}^{\gamma} \right ] },
\end{equation}
where $t_p = GM / {(\Delta v + \Omega R_H)}^3$ is the characteristic passing time scale with $\chi = 0.4$ and $\gamma = 0.65$ \citep{2010A&A...520A..43O}.

The planetary embryos grow by accreting pebbles until they reach the so-called pebble isolation mass, that is the mass required to perturb the gas pressure gradient in the disc: i.e. the gas velocity becomes super-Keplerian in a narrow ring outside of a planet's orbit reversing the action of the gas drag. The pebbles are therefore pushed outwards rather than inwards and accumulate at the outer edge of this ring stopping the embryos from accreting solids \citep{PaardekooperMellema06,Rice06}. Initial work found that the pebble isolation mass was proportional to the cube of the local gas aspect ratio \citep{2014A&A...572A.107L}. More recent work however has examined what effects disc viscosity and the Stokes number of the pebbles have on the pebble isolation mass, finding that small pebbles that are well coupled to the gas are able to drift past the pressure bump exterior to the planet's orbit \citep{Ataiee18,Bitsch18}. To account for the pebble isolation mass whilst including the effects of turbulence and stokes number, we follow \citet{Bitsch18}, and define a pebble isolation mass-to-star ratio,
\begin{equation}
q_{\rm iso} = \left(q_{\rm iso}^{\dagger} + \frac{\Pi_{\rm crit}}{\lambda} \right) \frac{M_{\oplus}}{M_*}
\end{equation}
where $q_{\rm iso}^\dagger  = 25 f_{\rm fit}$, $\lambda = 0.00476/f_{\rm fit}$, $\Pi_{\rm crit} = \frac{\alpha}{2St}$ and
\begin{equation}
\label{eq:ffit}
 f_{\rm fit} = \left[\frac{H/r}{0.05}\right]^3 \left[ 0.34 \left(\frac{\log(\alpha_3)}{\log(\alpha)}\right)^4 + 0.66 \right] \left[1-\frac{\frac{\partial\ln P}{\partial\ln r } +2.5}{6} \right] \ ,
\end{equation}
with $\alpha_3 = 0.001$.
Once planetary embryos reach the pebble isolation mass, they no longer accrete pebbles from the discs in our simulations.


\subsection{Gas envelope accretion}
\label{sub_sect_gasacc}
Once a planet reaches sufficient mass through pebble accretion, it is able to accrete a gaseous envelope from the surrounding disc.
Ideally we would incorporate 1D envelope structure models \citep[e.g.][]{CPN17} into our simulations.
However these calculations are computationally expensive and would considerably increase the simulation run times, therefore we opted to include gas accretion fits instead.
Previous work \citep{ColemanNelson16} provided fits to the 1D calculations presented in \citet{Movs}, that was for a planet located at 5.2 $\au$.
More recently, \citet{Poon21} presented an updated gas accretion model based on fits to gas accretion rates obtained using a 1D envelope structure model \citep{Pap-Terquem-envelopes,PapNelson2005,CPN17}.
To calculate these fits, \citet{Poon21} performed numerous simulations, embedding planets with initial core masses between 2--15 $\me$ at orbital radii spanning 0.2--50 $\au$, within gas discs of different masses.
This allowed for the effects of varying local disc properties to be taken into account when calculating gas accretion fits, a significant improvement on fits from other works \citep[e.g.][]{Hellary,ColemanNelson16}.
Using the 1D envelope structure model of \citet{CPN17}, the embedded planets were then able to accrete gas from the surrounding gas disc until either the protoplanetary disc dispersed, or the planets reached a critical state where they would then undergo runaway gas accretion.
With the results of these growing planets, \citet{Poon21} calculated fits to the gas accretion rates taking into account both the planet and the local disc properties.

Following the fits found in \citet{Poon21}, the gas accretion rate we use is equal to
\begin{align}\label{eq:gasenvelope-gc}
\left (\dfrac{d M_{\mathrm{ge}}}{d t} \right )_{\mathrm{local}}=& 10^{-10.199} \left( \dfrac{\mathrm{M_{\oplus}}}{\mathrm{yr}}\right) f_{\mathrm{opa}}^{-0.963} \left ( \dfrac{T_{\mathrm{local}}}{\mathrm{1\,K}}\right )^{-0.7049}\nonumber \\
&\times \left (\dfrac{M_{\mathrm{core}}}{\mathrm{M}_{\oplus}} \right )^{5.6549}  \left (\dfrac{M_{\mathrm{ge}}}{\mathrm{M}_{\oplus}}  \right )^{-1.159}\nonumber \\
&\times\left [ \exp{\left ( \dfrac{M_{\mathrm{ge}}}{M_{\mathrm{core}}} \right )} \right ]^{3.6334}.
\end{align}
where $T_{\rm local}$ is the local disc temperature, $f_{\rm opa}$ is an opacity reduction factor (which reduces the grain opacity contribution, kept constant at $f_{\rm opa} = 0.01$ in all simulations) and $M_{\rm core}$ and $M_{\rm ge}$ are the planet's core and envelope masses respectively.
When comparing the masses of gas accreting planets calculated through eq. \ref{eq:gasenvelope-gc} to the actual masses obtained using the 1D envelope structure model of \citet{CPN17}, \citet{Poon21} found excellent agreement, a considerable improvement on previous fits \citep[e.g.][]{ColemanNelson16}.
In each of our simulations, we allow the planetary embryo to accrete a gaseous envelope once their mass exceeds an Earth mass, and all mass accreted by planets is removed from the disc to conserve mass.

\subsection{Type I migration}
\label{subsect_typeImig}
In our simulations, planetary embryos that grow massive enough (significantly exceeding a Lunar mass) are subject to type I migration, which is implemented in the same way as \citet{2014MNRAS.445..479C}, based on the torque formulae presented by \citet{2010MNRAS.401.1950P,2011MNRAS.410..293P}. These formulae specify how the different torques (that contribute to the total torque acting on the planetary embryo, expressed by equation \ref{eqn:typeItorque}) change with planet mass and local disc conditions (determined by changes in temperature, surface density and metallicity/opacity). Corotation torques in particular vary sensitively with the ratio of the horseshoe libration time-scale to either the viscous or thermal diffusion time-scales. 

The torque experienced by a low mass planet embedded in a disc is contributed by the Lindblad torques and a weighted sum of the vorticity-related horseshoe drag, the entropy related horseshoe drag, the vorticity-related linear corotation torque, and the entropy-related linear corotation torque. Combining equations 50-53 in \citep{2011MNRAS.410..293P}, we express the total torque for type I migration $\Gamma_{\textrm{tot}}$ acting on a planet as:

\begin{multline}
\label{eqn:typeItorque}
    \Gamma_{\textrm{tot}} = \Gamma_{LR} + \{ \Gamma_{\textrm{VHS}} F_{\nu} G_{\nu} + \Gamma_{\textrm{EHS}} F_{\nu} F_{d} \sqrt{G_{\nu} G_{d}}  \\ + \Gamma_{\textrm{LVCT}}(1-K_{\nu}) +\Gamma_{\textrm{LECT}}(1-K_{\nu})(1-K_d)\}, 
\end{multline}
where $\Gamma_{\textrm{LR}}$ is the Lindblad torque,  $\Gamma_{\textrm{VHS}}$ is the vorticity-related horseshoe drag, $\Gamma_{\textrm{EHS}}$ is the entropy-related horseshoe drag, $\Gamma_{\textrm{LVCT}}$ is the vorticity-related linear corotation torque, $\Gamma_{\textrm{LECT}}$ is the entropy-related linear corotation torque \citep[as given by equations 3-7 in][]{2011MNRAS.410..293P}. The functions $G_{\nu}$, $G_d$, $F_{\nu}$, $F_d$, $K_{\nu}$, $K_d$ are related to either the ratio between the viscous/thermal diffusion time scales and the horseshoe liberation time-scale, or to the ratio of the viscous/thermal diffusion time-scales and the horseshoe U-turn time-scale \citep[see equations 23, 30 and 31 in][]{2011MNRAS.410..293P}. As our simulations only consider circular coplanar orbits, the effects of a planet's eccentricity and inclination on attenuating the Lindblad and corotation torques are not required, although these components are present in the code \citep[see][]{2014MNRAS.445..479C}). \footnote{Note that although our code includes the component for type II migration, none of the planets in our simulations grew massive enough to open a gap in the disc and switch to type II migration, according to the gap opening criterion in \citet{2006Icar..181..587C}.}

\section{Results and discussion}

We begin by studying the mass budget and duration of pebble flux through the disc as a function of external FUV field strength and shielding time-scale in \ref{subsec:t_cutoff}. We then move on to discuss the impact on the planets forming in our models in section \ref{subsec:planetGrowthMigration}. Finally in \ref{sec:boundedness} we discuss our understanding of the shielding time, how that links to the star formation process and how that affects our ability to predict exoplanet populations.

\subsection{Pebble cutoff time}
\label{subsec:t_cutoff}
Planet formation by pebble accretion relies on a flux of pebbles from the outer disc. Grains undergo radial drift once they reach sufficient size to dynamically decouple from the gas \citep{1977MNRAS.180...57W}. The time-scale for growth to such a size increases radially outwards through the disc \citep[e.g.][]{2012A&A...539A.148B}, resulting in a ``pebble production front'' at radius $r_g$. Once the pebbles have grown to a size where they undergo radial drift they are effectively safe from external photoevaporation, both because only small grains are entrained in a wind \citep{2016MNRAS.457.3593F} and because radial drift moves them in to smaller radii where external photoevaporation does not operate.

The \textit{pebble cut-off time}, $t_{\textrm{cutoff}}$, is the time a disc has before its pebble production front $r_g$ reaches the disc outer edge, cutting off the supply of pebbles. $t_{\textrm{cutoff}}$ relative to the time that the pebble production front passed a planetary embryo's location determines the time available for the embryo to grow via pebble accretion. 
If a disc is subject to external photoevaporation and the mass loss rate is greater than the rate of viscous spreading the disc will be truncated \citep{2007MNRAS.376.1350C}. Consequently, the disc outer edge moves inwards and meets the outward-moving pebble production front earlier than it otherwise would have. This reduces the pebble cut-off time compared to when the disc is not truncated. Therefore how much the pebble cut-off time is shortened, i.e. how long the planetary embryos have to accrete pebbles, depends on the strength of the radiation field and the shielding time during which the disc is protected from truncation via external photoevaporation. 

\begin{figure}
    \centering	\includegraphics[width=\columnwidth]{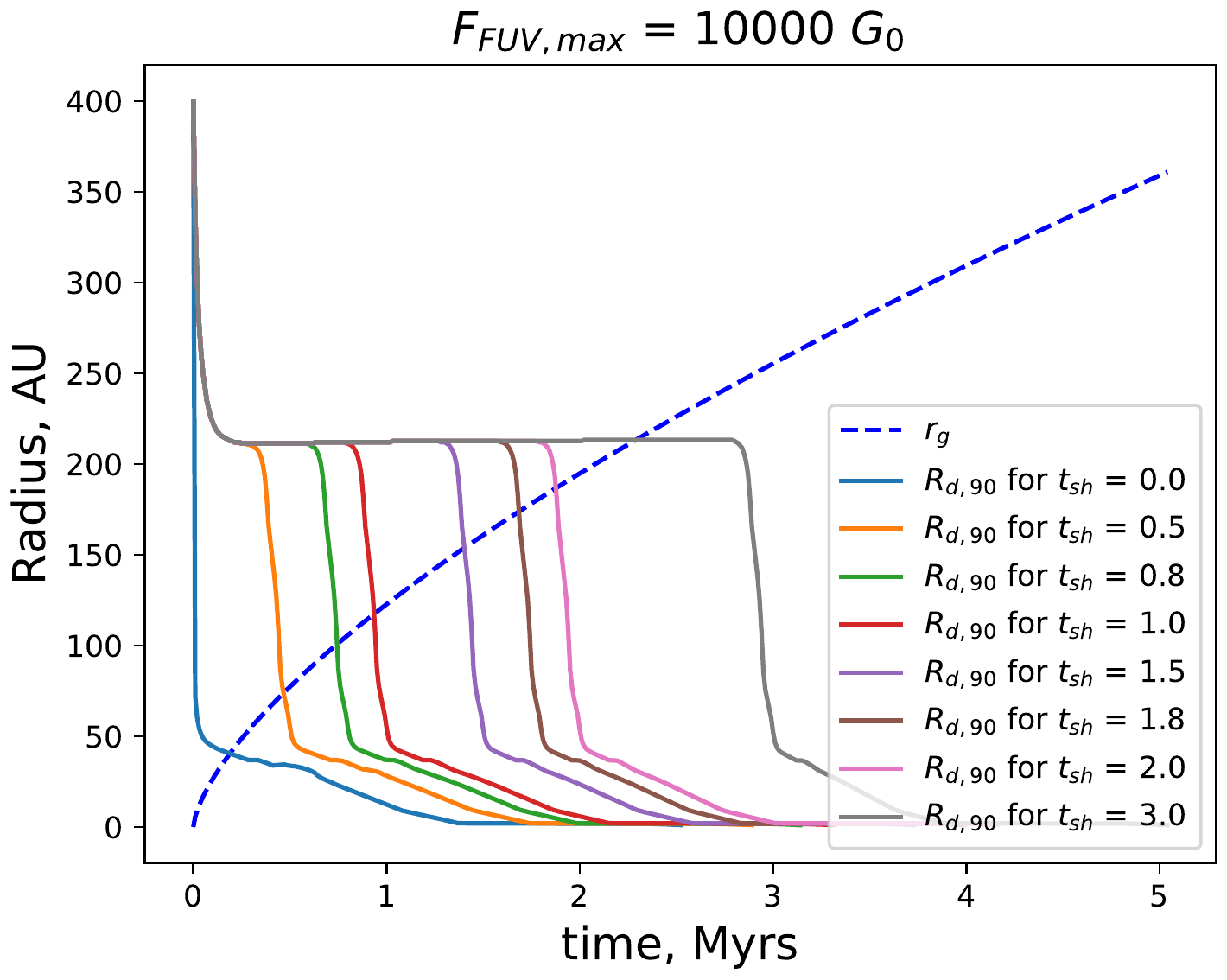}
    \caption{The pebble production front $r_g$ as a function of time for the disc configuration simulated (dashed curve) and disc radii $R_{\textrm{d,90}}$ evolution under and FUV track with $F_{\textrm{FUV, max}} =10^4$ $G_0$ and various shielding times (solid lines). Longer shielding means that $r_g$ intersects the disc outer radius at a later time, resulting in a longer period of pebble flux through the disc.} 
    \label{fig:radius_rpebble_10000G0}
\end{figure}

The dashed line in Figure \ref{fig:radius_rpebble_10000G0} shows the pebble production front $r_g$ over time in our modelled disc. It compares the production front radius with the disc outer edge when subjected to a strong FUV-irradiation field($F_{\textrm{FUV, max}} = 10^4 G_0$), for various shielding times $t_{\textrm{sh}}$. Note that here we define the disc outer edge (or disc radius) as the radius which contains 90\% of the gas mass, which we denote $R_{\textrm{d,90}}$. It shows that as soon as the disc is exposed to the strong FUV field, the disc radius is rapidly truncated down to $\sim$40\,au, then undergoes a much slower phase of truncation until fully dispersed. A longer shielding time protects the disc from the rapid truncation phase and hence the pebble cut-off time, i.e. the time when $r_g$ crosses $R_{\textrm{d,90}}$ gets longer as $t_{\textrm{sh}}$ increases. 

\begin{figure}
    \centering	\includegraphics[width=\columnwidth]{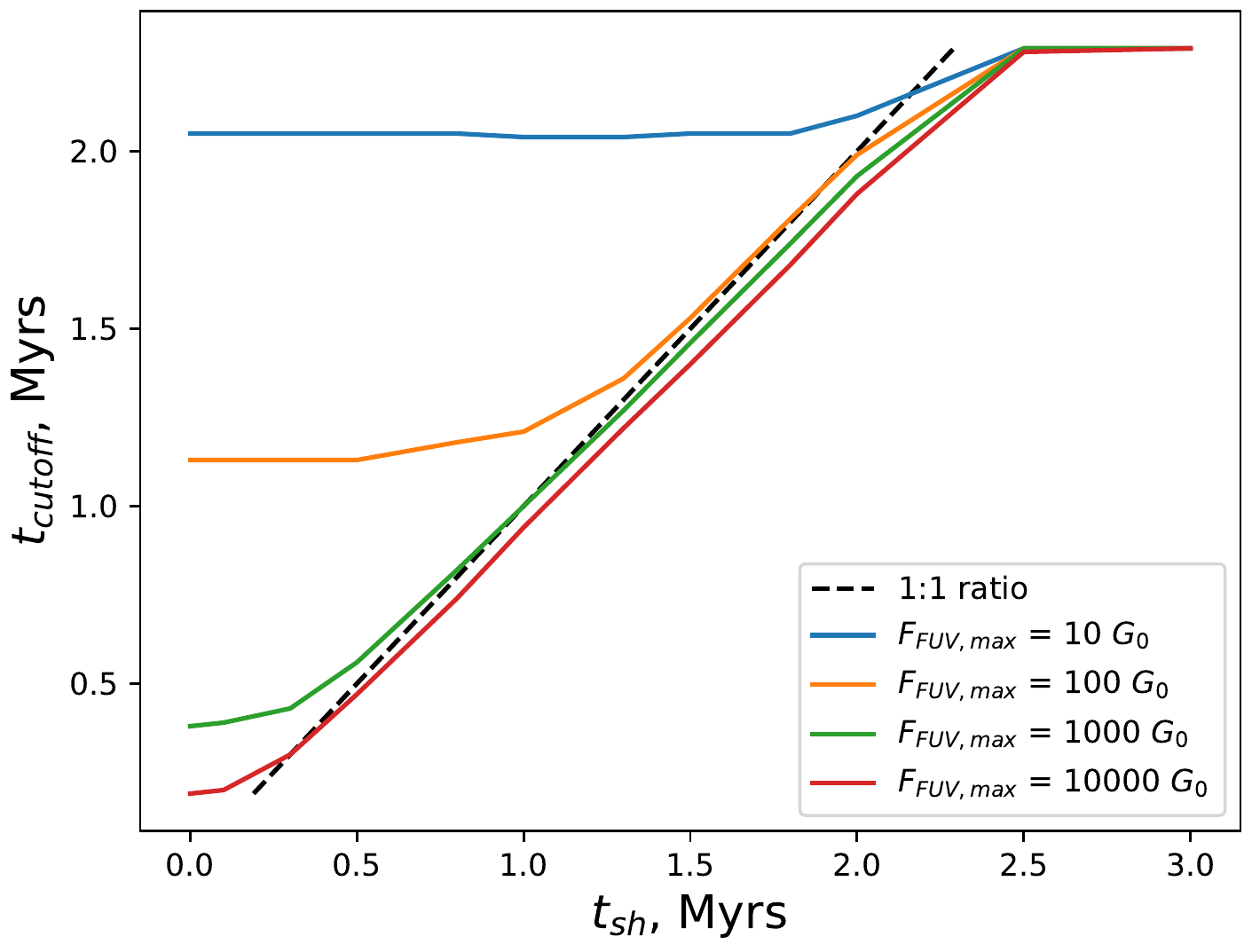}
    \caption{The pebble cut-off time $t_{\textrm{cutoff}}$ as a function of the shielding time $t_{\textrm{sh}}$ (solid lines) for FUV tracks of different $F_{\textrm{FUV, max}}$ values indicated by different colors. The dashed line highlights the 1:1 ratio as a reference.} 
    \label{fig:pebbles_cutofftime}
\end{figure}

The correlation of pebble cut-off time with shielding time depends on $F_{\textrm{FUV, max}}$, as illustrated in  Figure \ref{fig:pebbles_cutofftime} which shows $t_{\textrm{cutoff}}$ as a function of $t_{\textrm{sh}}$ for discs evolved under various $F_{\textrm{FUV, max}}$. For the disc evolved with $F_{\textrm{FUV, max}} = 10^4$ 
 $G_{0}$ (the strongest FUV strength in figure \ref{fig:radius_rpebble_10000G0}), $t_{\textrm{cutoff}}$ and $t_{\textrm{sh}}$ are almost in a 1:1 relationship for $t_{\textrm{sh}} < 2.5$ Myrs. This is due to the disc outer edge being immediately stripped to inside of $r_g$ as soon as the shielding ends. By 2.5 Myrs, $r_g$ reaches the original disc outer edge, resulting in the maximum possible cutoff time, and so longer shielding beyond 2.5 Myrs has no additional impact. For lower values of $F_{\textrm{FUV, max}}$ in the range of $t_{\textrm{sh}} < 2.5$ Myrs, $t_{\textrm{cutoff}}$ initially stays almost constant for shorter lengths of $t_{\textrm{sh}}$ (because weaker FUV radiation reduces the disc size less efficiently), before again becoming linearly correlated to $t_{\textrm{sh}}$. The weaker the $F_{\textrm{FUV, max}}$, the longer the $t_{\textrm{cutoff}}$ remains constant before starting to increase with $t_{\textrm{sh}}$. 
 
 Overall fig. \ref{fig:pebbles_cutofftime} shows that the cloud shielding time is important in determining $t_{\textrm{cutoff}}$ over a wide range of UV field strengths. This affects the total mass budget in drifting pebbles and the time-scale over which there is a pebble flux through the disc. We next turn our attention to how this affects planet formation by pebble accretion.

\subsection{Planet growth and migration}
\label{subsec:planetGrowthMigration}

A planetary embryo begins accreting pebbles once the pebble production front reaches the orbital radius of the embryo ($r_g=a$). This continues until shortly after the pebble growth front reaches the disc outer edge (i.e. at $t_{\textrm{cutoff}}$). This in turn depends on the FUV radiation strength and shielding time, since those control the evolution of the disc outer radius.

\begin{figure*}
    \centering
    \includegraphics[width=2\columnwidth]{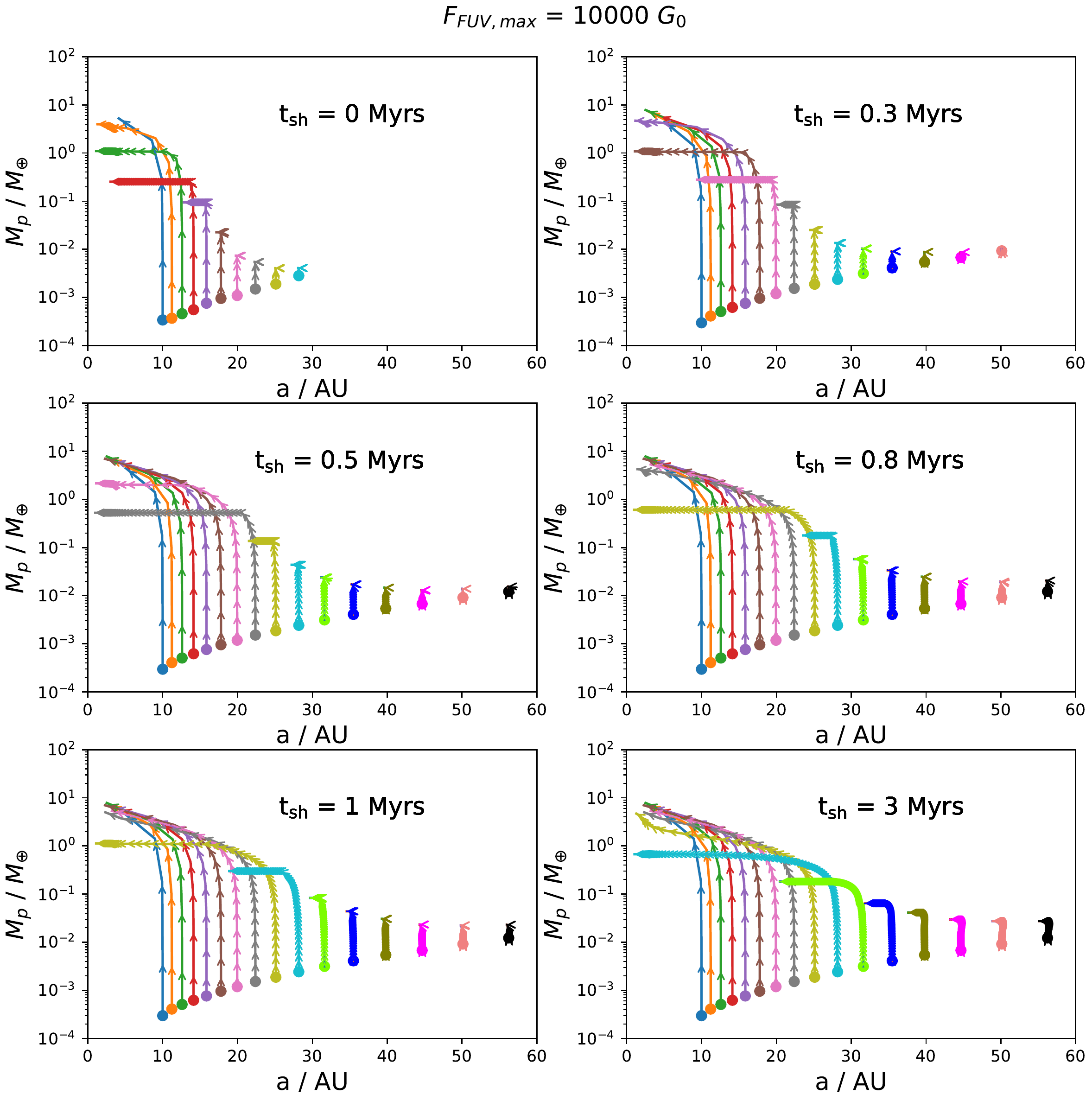}
    \caption{Evolution of planetary embryo mass $M_{p}$ against semi-major axis $a$ in discs evolved with FUV tracks with $F_{\textrm{FUV, max}}$ = $10^4$ $G_0$. In each subplot, simulations with one specific shielding time $t_{\textrm{sh}}$ are included, and each colored curve represents the evolution trajectory of one planetary embryo in one disc with arrows indicating the direction of evolution. The initial planetary embryo core mass and semi-major axis ($a_{\textrm{init}}$) when injected is indicated by the circle. Note that the plots only includes the simulations with $a_{\textrm{init}} < 60$ AU though the full $a_{\textrm{init}}$ range simulated is 10 - 100 AU.}
    \label{fig:m_aplot}
\end{figure*}

Figure \ref{fig:m_aplot} shows the evolution of planet masses $M_{p}$ and semi-major axes $a$, for discs irradiated by a $10^4$\,$G_0$ radiation field after some shielding time. The circles represent the initial mass/semi-major axis of the planetary embryos, and the arrowed tracks show the subsequent evolution. Each panel represents the evolution of sets of planetary embryos for different shielding time-scales, from no shielding in the upper left, to 3\,Myr of shielding in the lower right panel. 

This figure illustrates how the cloud shielding time affects the growth and migration of planetary embryos at different semi-major axes.  Firstly the top-left subplot with $t_{\textrm{sh}} = 0$ demonstrates the effects of strong external photoevaporation in suppressing the core growth by the fast truncation of the disc outer edge and hence quickly cutting off the supply of pebble flux (shortening $t_{\textrm{cutoff}}$) for planetary embryo cores injected at almost all semi-major axes. For $r >$ 35 au, the disc is truncated so efficiently that the disc outer edge has already moved inside the pebble production front $r_g$, before $r_g$ reaches 35 au, and so no embryo cores were injected beyond 35 au. Even within 35 au, there is hardly any growth and migration for cores injected at > 20 au due to short $t_{\textrm{cutoff}}$, and only the innermost three cores (injected at 10, 11 and 13 au) had enough time to grow and accrete meagre gaseous envelopes but only reached masses between 1-10 $M_{\oplus}$.

\begin{figure}
    \centering	\includegraphics[width=\columnwidth]{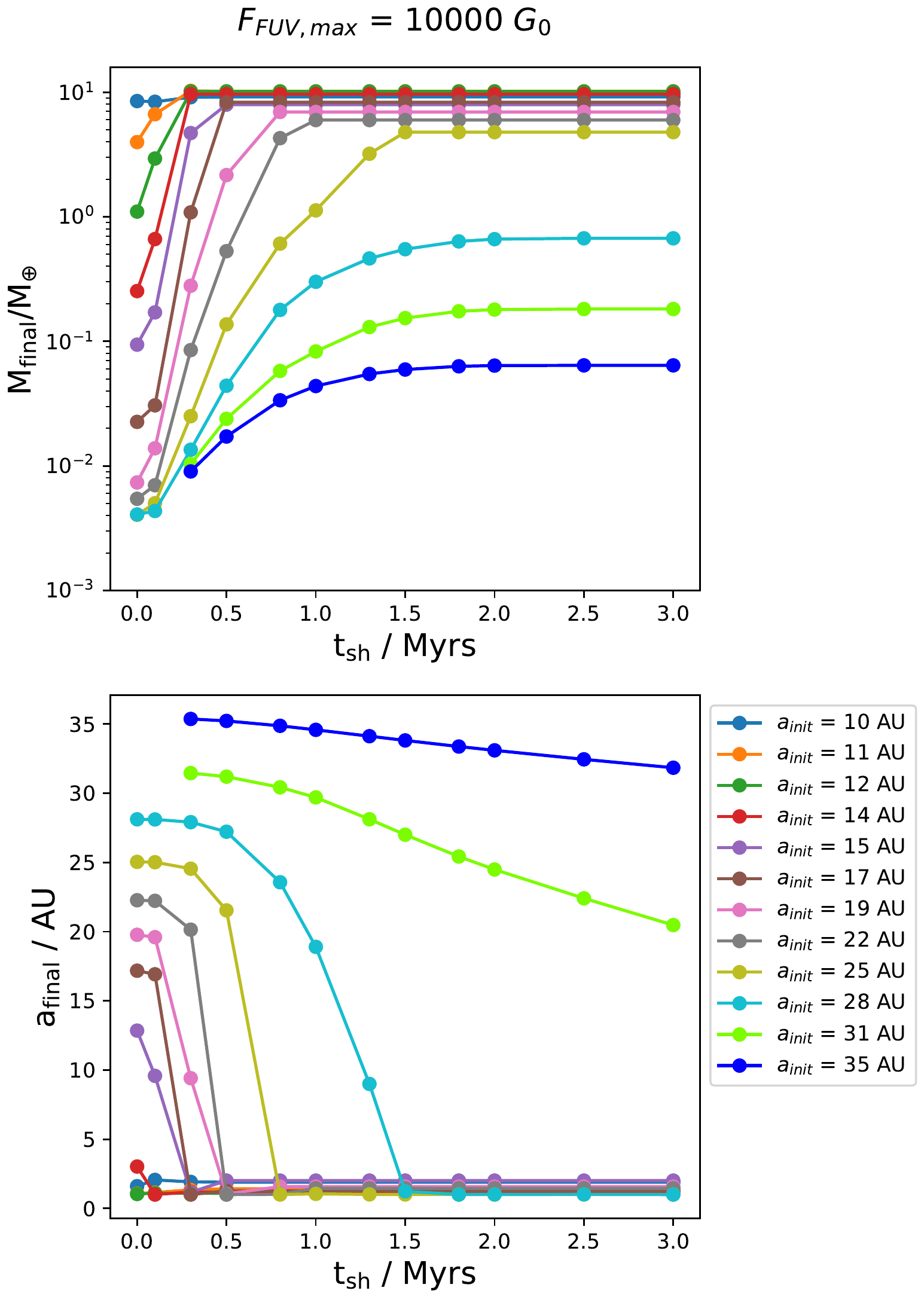}
    \caption{The final mass $M_{\textrm{final}}$ (upper panel) and final semi-major axis $a_{\textrm{final}}$ (lower panel) that the injected planetary embryos reached as a function of shielding time $t_{\textrm{sh}}$ in discs evolved with the FUV radiation track of $F_{\textrm{FUV, max}} =$ $10^4$ $G_{0}$, with different colors indicating the initial semi-major axis $a_{\textrm{init}}$ of the planetary embryos.} 
    \label{fig:mfinal_tsh_10000G0}
\end{figure}

 With such strong irradiation, even a short shielding time makes a difference in delaying rapid truncation and increase $t_{\textrm{cutoff}}$, allowing the cores more time to receive the pebble flux and grow. The top right subplot with $t_{\textrm{sh}} = 0.3$ Myrs, for example, already shows a difference in the final masses that the cores were able to reach, especially for those with $a_{\textrm{init}}$ between 14 and 25\,au, whose final masses are at least one order magnitude larger than the case with $t_{\textrm{sh}} = 0$. As $t_{\textrm{sh}}$ increases, more cores injected at larger $a_{\textrm{init}}$ are able to accrete for longer and so grow more massive and migrate. This behaviour is demonstrated further in the top plot of figure \ref{fig:mfinal_tsh_10000G0}, which plots the final masses $M_{\textrm{final}}$ that the planetary embryos injected at various $a_{\textrm{init}}$ were able to reach as a function of $t_{\textrm{sh}}$.

The top plot of figure \ref{fig:mfinal_tsh_10000G0} shows that for a planetary embryo injected at some radial distance $a_{\textrm{init}}$, the $M_{\textrm{final}}$ it is able to reach increases with shielding time initially until a certain value of $t_{\textrm{sh}}$, after which $M_{\textrm{final}}$ remains roughly the same. This is due to the shielding time being long enough for the pebble production front to reach the disc outer edge before being exposed to external photoevaporation. In that case the disc retains the maximal pebble mass budget and so extending the shielding time-scale further has no impact on planet formation. 
 
The shielding time makes the most difference for the embryos injected between $a_{\textrm{init}}$ = 14 and 31 AU in terms of the steepness of the correlation between $M_{\textrm{final}}$ and $t_{\textrm{sh}}$ before plateauing. Embryos in this radial range reach masses of around a lunar mass ($\sim 0.01 M_{\oplus}$) when not shielded, but were able to reach $\sim 5 M_{\oplus}$ when shielded for longer than $\sim 1.5$ Myrs. With long enough $t_{\textrm{sh}}$, all the embryos with $a_{\textrm{init}} <=$ 31 AU are able to reach at least 0.1 $M_{\oplus}$, with those of $a_{\textrm{init}} <=$ 25 AU able to achieve $1 - 10$ $M_{\oplus}$. For the embryos injected with $a_{\textrm{init}} >$ 31 AU, as it takes longer for $r_g$ to reach their location so that they can start accreting pebbles, the time left for their pebble accretion before $r_g$ crosses disc outer edge is not long enough for them to grow larger than 0.1 $M_{\oplus}$ even with relatively long $t_{\textrm{sh}}$. The opposite occurs for the embryos with $a_{\textrm{init}} <$ 14 AU, as $r_g$ reaches their locations quite early on, they had enough time to grow beyond 1 $M_{\oplus}$ and migrate inward rapidly even without shielding.

\begin{figure}
    \centering	\includegraphics[width=\columnwidth]{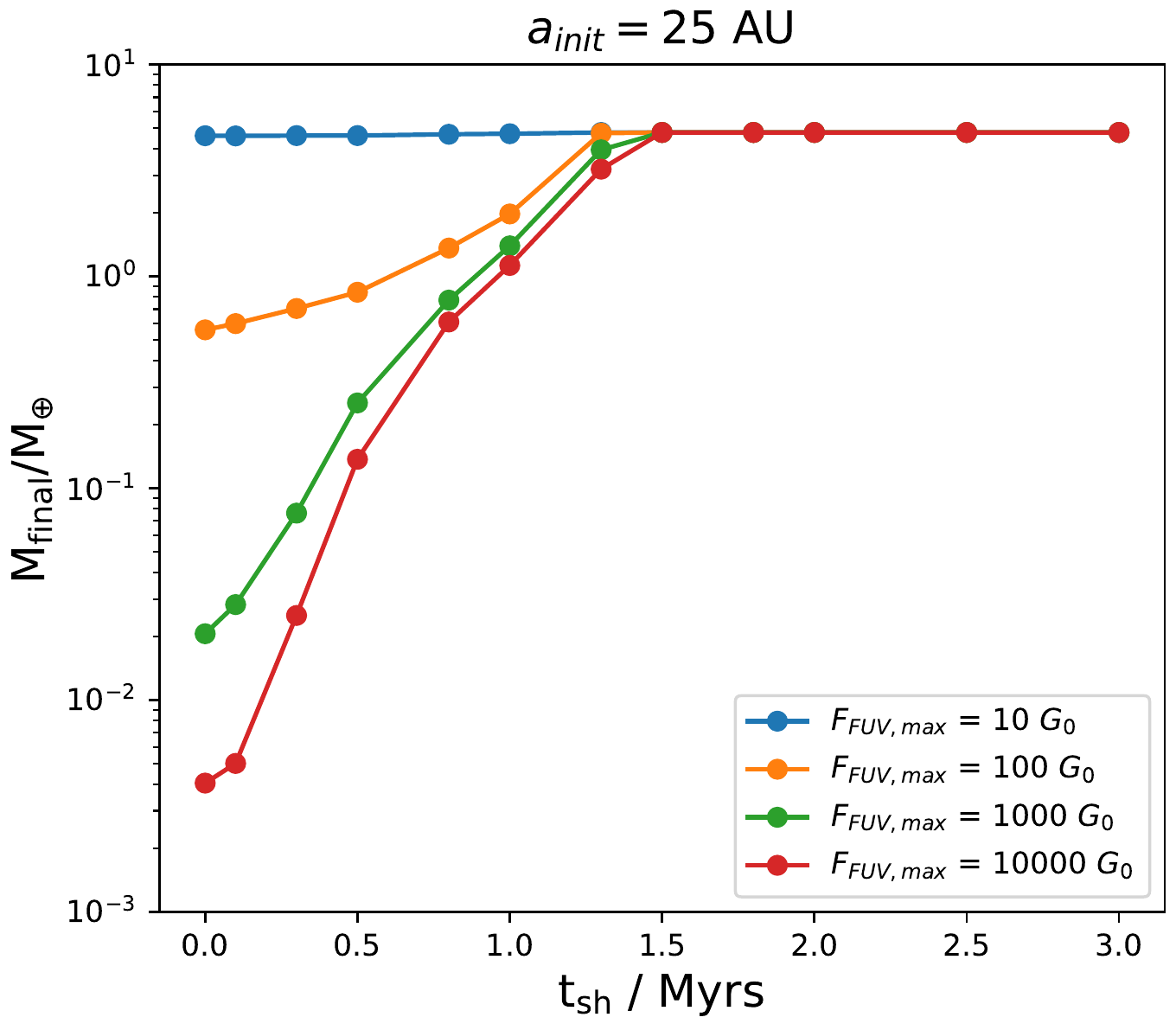}
    \caption{The final mass $M_{\textrm{final}}$ that the planetary embryos injected at $a_{\textrm{init}} =$ 25 AU reached as a function of shielding time $t_{\textrm{sh}}$ in discs evolved with FUV tracks of different $F_{\textrm{FUV, max}}$ values, indicated by different colors. There is a nonlinear impact of shielding time on the planet's final mass up to 1\,Myr for $F_{\textrm{FUV, max}}\geq100$\,G$_0$.}
    \label{fig:mfinal_tsh_ainit25}
\end{figure}

\begin{figure}
    \centering	\includegraphics[width=\columnwidth]{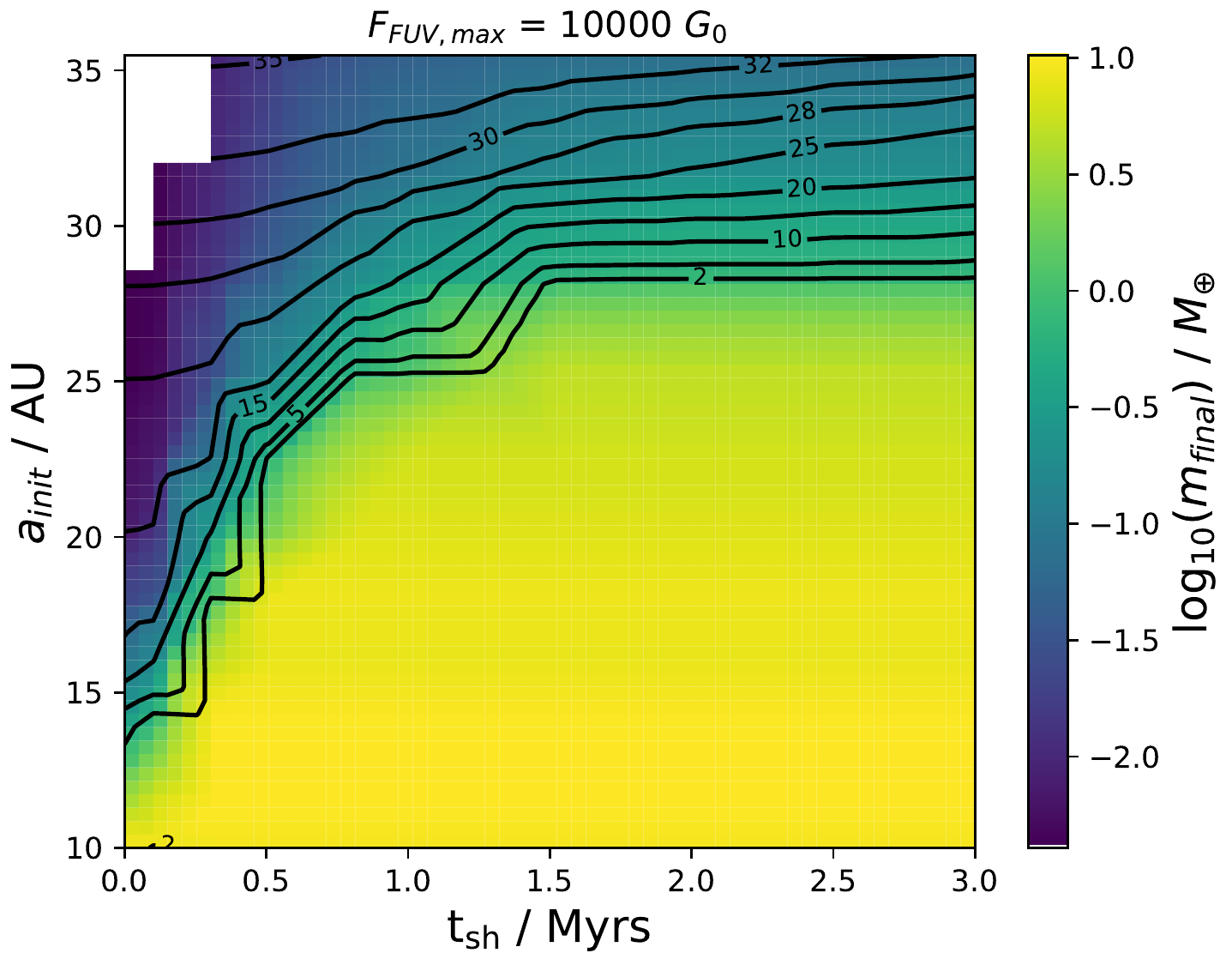}
    \caption{The colour scale shows the final mass of a planet as a function of the shielding time and initial orbital radius of the planetary seed, in a $10^4$\,G$_0$ environment. The contours show the final semi-major axis ($a_{\textrm{final}}$) of the planets in au.  }
    \label{fig:contourPlot}
\end{figure}

Figures \ref{fig:m_aplot} and \ref{fig:mfinal_tsh_10000G0} demonstrate that the shielding time makes a significant difference to the growth and migration of planets in $10^4$\,G$_0$ environments. In section \ref{subsec:t_cutoff} we showed that $t_{\textrm{sh}}$ also affects $t_{\textrm{cutoff}}$ in lower FUV radiation environments. To explore this, Figure \ref{fig:mfinal_tsh_ainit25} shows how the final mass of an object resulting from an embryo injected at $a_{\textrm{init}} =$ 25\,au scales as a function of the FUV radiation environment once the disc is unshielded. The final planet mass scales non-linearly with  $t_{\textrm{sh}}$ up to around 1.5\,Myr in the high UV field cases ($10^3-10^4$\,G$_0$).  Even in the case of a 100\,G$_0$ radiation field the difference in mass between no shielding and 1.5\,Myr of shielding is an order of magnitude.  Figure \ref{fig:contourPlot} also shows a contour plot that summarises how the final planet mass/semi-major axes varies as a function of the initial semi-major axis and shielding time-scale. 


Beyond a certain shielding time-scale ($\sim1.5\,$Myr in this case) the mass reservoir for pebble accretion has reached the disc outer edge and further shielding from external photoevaporation has no additional impact on planet properties. Beyond such a time, other processes, such as planet migration and N-body interactions will have a greater impact on the types of planets that form in the simulations.



\subsection{The boundedness of clouds and shielding time-scales of discs}
\label{sec:boundedness}
We have demonstrated that planet formation by pebble accretion, a widely used planet formation paradigm, is sensitive to external photoevaporation. Furthermore, it is sensitive to the time-scale that a disc can be shielded from external photoevaporation up to shielding time-scales of $\sim1.5$\,Myr. We now turn our attention to the expected shielding time-scales of discs. 

It is well established in numerical simulations that the extent and rapidity with which a star forming cloud is dispersed by feedback is a function of the cloud boundedness (virial ratio), see for example \cite{2012MNRAS.424..377D, 2012MNRAS.427.2852D, 2013MNRAS.430..234D, 2013MNRAS.431.1062D}, whose simulations demonstrate that globally bound clouds are dispersed more slowly. Since the rate that the molecular gas is dispersed decreases for more bound clouds, we therefore expect that the shielding time-scale will also correlate with the boundedness of the cloud. Other recent studies of feedback in clouds find that protostellar jets and other forms of feedback can play a role in supporting against very low virial ratios, preventing it from dropping significantly below unity on a global scale of the cloud \citep{2022MNRAS.512..216G, 2022MNRAS.515.4929G}

The expectation that shielding will be affected by cloud boundedness is supported by recent simulations that study the expected UV irradiation of discs in star formation simulations \citep{2022MNRAS.512.3788Q, 2023arXiv230203721W}. These papers studied star formation from clouds with initial global virial ratios of $\alpha=2$ and 0.25 respectively, which effectively correspond to an upper limit on the virial ratio for collapse to occur and lower extreme of observed virial ratio respectively.  In accordance with our expectations from the discussion above, \cite{2022MNRAS.512.3788Q}  and \cite{2023arXiv230203721W} find shielding time-scales of $<0.5$\,Myr and of \textit{at least} $0.5$\,Myr respectively. There will be a continuum of behaviours between these two examples, which is yet to be explored in detail. It will be important to do so, since  when coupled with our results, this range of shielding times could either lead to planet formation being dramatically influenced by the environment for a shielding time $<0.5$\,Myr, to it being a much more marginal (or even negligible) effect for shielding $>1.5\,$Myr.

We note that measured virial ratios of clouds vary broadly \citep[see e.g. Figure 1 of][where they vary from $\alpha\sim3\times10^{-2}$ to almost $10^3$]{2013ApJ...779..185K}. A large marginally bound cloud like that of \cite{2022MNRAS.512.3788Q} will end up with very sub-virial clumps at the sites of massive star formation \citep[see e.g.][for simulations following the virial ratio of substructures from larger scale turbulent clouds]{2020ApJ...903...46C}.  We also note that the virial ratio is also easy to be underestimated through systematics, for example as highlighted by \cite{2021ApJ...922...87S} for Gould Belt clumps \cite[see also][]{2018A&A...619L...7T}

The discovery of proplyds in the very young (0.2-1\,Myr) star forming region NGC 2024 demonstrates that external photoevaporation certainly does happen at a time that would influence planet formation by pebble accretion \citep[according to our models, ][]{2021MNRAS.501.3502H}. At the other extreme \cite{2022A&A...662A..74G} observed a disc with signs of planet formation emerging from an irradiated cloud in a region that is $\sim\,5$Myr, raising the possibility of a very long shielding time-scale. Observationally then, it is clear that both very short and long shielding time-scales are happening, but the actual probability distribution for real systems across the wide range of star forming scenarios is yet to be established theoretically or observationally.

\subsection{Implications for planet populations}
We have demonstrated here that planet formation by pebble accretion can be affected by the radiation environment, and that the shielding time-scale can have a nonlinear impact on the resulting planets. However the uncertainty of the shielding time-scale distribution and the likely coupling of that distribution to the particular star formation scenario could provide a major source of uncertainty in planetary population synthesis models. 

Whilst we have shown that the shielding time-scales can have non-linear effects on the resulting planets, our results also show that the radiation environment itself also has a large effect. Indeed, from fig. \ref{fig:mfinal_tsh_ainit25} when there is little shielding, there is a large difference in the final planet masses for those forming at 25$\au$ in weak environments (e.g. 10G$_0$) to those in strong environments (e.g. 10$^4$ G$_0$). The impact on the planetary population is therefore likely to vary from cluster to cluster. In the strong UV environments of massive star forming regions, such as NGC 2024, the ONC and Carina \citep{2020MNRAS.491..903W} this could indicate that the final planet population might be significantly different to that which would form in a weak environment, e.g. Taurus, mainly due to the reduction in pebbles and solids that are able to be accreted, as shown above. The differences in the environment will also affect the scale of migration that planets can undergo, that will influence the final populations that can be observed, as both seen here and in \cite{2022MNRAS.515.4287W}.

There could also be additional processes that are affected by the early truncation of the protoplanetary disc through the removal of shielding from the parent cloud. For example a significant simplification of our model is that we do not consider the effect of substructures in the disc on the dust evolution. These can cause pressure bumps that control the distribution of dust in a disc. 
Here we assumed that the cores of planets can only grow through the accretion of pebbles.
However, recent work that examined the population of planetesimals and planetary embryos that arise through the gravitational collapse of pebbles trapped at specific locations in the disc \citep{Lenz19,2021MNRAS.506.3596C}, could also be affected by the efficient truncation of the disc.
Since the formation of planetesimals, and subsequently planetary embryos, may be dependent on the pebble production front, it is reasonable to assume that the effects on the front discussed in this paper will also affect their formation as well.
This will further reduce the amount of solids available for accretion by planets, whilst simultaneously reducing the number and/or mass of the initial planetary embryos.
With fewer, less massive planets, this could result in large discrepancies in the populations from one cluster to another, depending purely on the amount of shielding, which we will explore in future work.

An additional effect that would affect planetesimal and embryo formation, is that the removal of gas would lead to higher dust-to-gas ratios in the outer disc. This may lead to an increase in the planetesimal formation rate via the streaming instability \citep{Carrera17}. This could address some of the issues around a reduced planetesimal population through the pebble production front reaching the disc outer edge.

\section{Summary and conclusions}
In this paper we explored how planet formation by pebble accretion is affected by external photoevaporation, and how the impact of external photoevaporation can be mitigated by shielding the disc by embedding it in a molecular cloud for some time. We use a detailed and well established disc evolution and pebble accretion model coupled with the FRIED grid of mass loss rates to implement external photoevaporation. We draw the following main conclusions from this work:
\begin{enumerate}
    \item It is already well established that a growth front of pebbles propagates from the inner disc to the disc outer edge. Those pebbles then undergo radial drift inwards, providing a mass flux for planet formation. External photoevaporation can truncate the outer disc dramatically (and rapidly). We show that if that happens before the pebble growth front reaches the disc outer edge, it acts to reduce the mass reservoir of pebbles and the time-scale over which there will be a pebble flux throughout the disc. Because the pebbles undergo radial drift, this impact on the outer disc can in principle affect planet formation anywhere in the disc. 
    \item The shielding time-scale has a nonlinear impact on the resulting planet mass for planetary embryos originating beyond about 14\,au. Note that planets formed originally at these locations in the disc do migrate in to short period orbits such as those observed for many exoplanet systems. A $10^{-3}\,M_\oplus$ planteary embryo at 25\,au, for example, stays at 25\,au with a lunar mass if the disc is immediately irradiated by a $10^3$\,G$_0$ field, but grows and migrates to be approximately Earth-like in both mass and orbital radius if the disc is shielded for just 1\,Myr. Even relatively brief periods of shielding ($<0.5$\,Myr) can drastically affect the final planetary system. Conversely, shielding for time-scales $>1$\,Myr essentially nullifies the impact of external photoevaporation on the resulting planets. 
    \item The distribution of shielding time-scales that all stars experience is not well understood, but does have some constraints. Feedback clears less bound clouds more rapidly, exposing discs to UV radiation on a shorter time-scale. \cite{2022MNRAS.512.3788Q} simulations of marginally bound clouds find that shielding is $<0.5$\,Myr, so a time-scale that is impactful for planet formation by pebble accretion according to our models here. Conversely at an extremely low global virial ratio \cite{2023arXiv230203721W} find shielding times of at least $0.5$\,Myr. Observed cloud virial ratios are uncertain, and span a range of values (discussed more in \ref{sec:boundedness}). Furthermore externally irradiated discs are observed both at very young (perhaps as young as 0.2\,Myr in NGC 2024) and much older ages. We hence do expect that planet formation by pebble accretion can be affected by the environment in practice. But how widespread this impact is remains to be determined, which provides significant uncertainty for planetary population synthesis models 
\end{enumerate}

In summary, we demonstrated that external photoevaporation and cloud shielding can have potentially significant impacts on planets formed via pebble accretion. Here we used a simplified model of one planet per disc and a parameterised FUV track with a constant field strength ($F_{\textrm{FUV,max}}$) after a shielding period. However, in a dynamically evolving star forming region, the FUV field radiated upon a disc constantly varies with time \citep{2022MNRAS.512.3788Q, 2023arXiv230203721W}. In order for a realistic study of to what extent the formation of planets are affected by the cluster they are born in, planet formation models should be coupled to the FUV tracks traced from stellar cluster and feedback simulations. Including multiple planets in one disc could result in higher mass cores via collisions, which could also act to enhance the pebble accretion rate. These more complicated models will be explored in future work.

\section*{Acknowledgements}
TJH is funded by a Royal Society Dorothy Hodgkin Fellowship. 
GALC was funded by the Leverhulme Trust through grant RPG-2018-418.
This work was performed using the DiRAC Data Intensive service at Leicester, operated by the University of Leicester
IT Services, which forms part of the STFC DiRAC HPC Facility (www.dirac.ac.uk). The equipment was funded by BEIS capital
funding via STFC capital grants ST/K000373/1 and ST/R002363/1
and STFC DiRAC Operations grant ST/R001014/1. DiRAC is part
of the National e-Infrastructure.
This research utilised Queen Mary's Apocrita HPC facility, supported by QMUL Research-IT (http://doi.org/10.5281/zenodo.438045). We thank the anonymous reviewer for their comments  which significantly improved the clarity of the paper..

\section*{Data Availability}
The simulation results are available upon request to the corresponding author.



\bibliographystyle{mnras}
\bibliography{Planet_form_paper} 




\appendix
\section{Examples of the disc surface density evolution}

\begin{figure}
    \centering
	\includegraphics[width=\columnwidth]{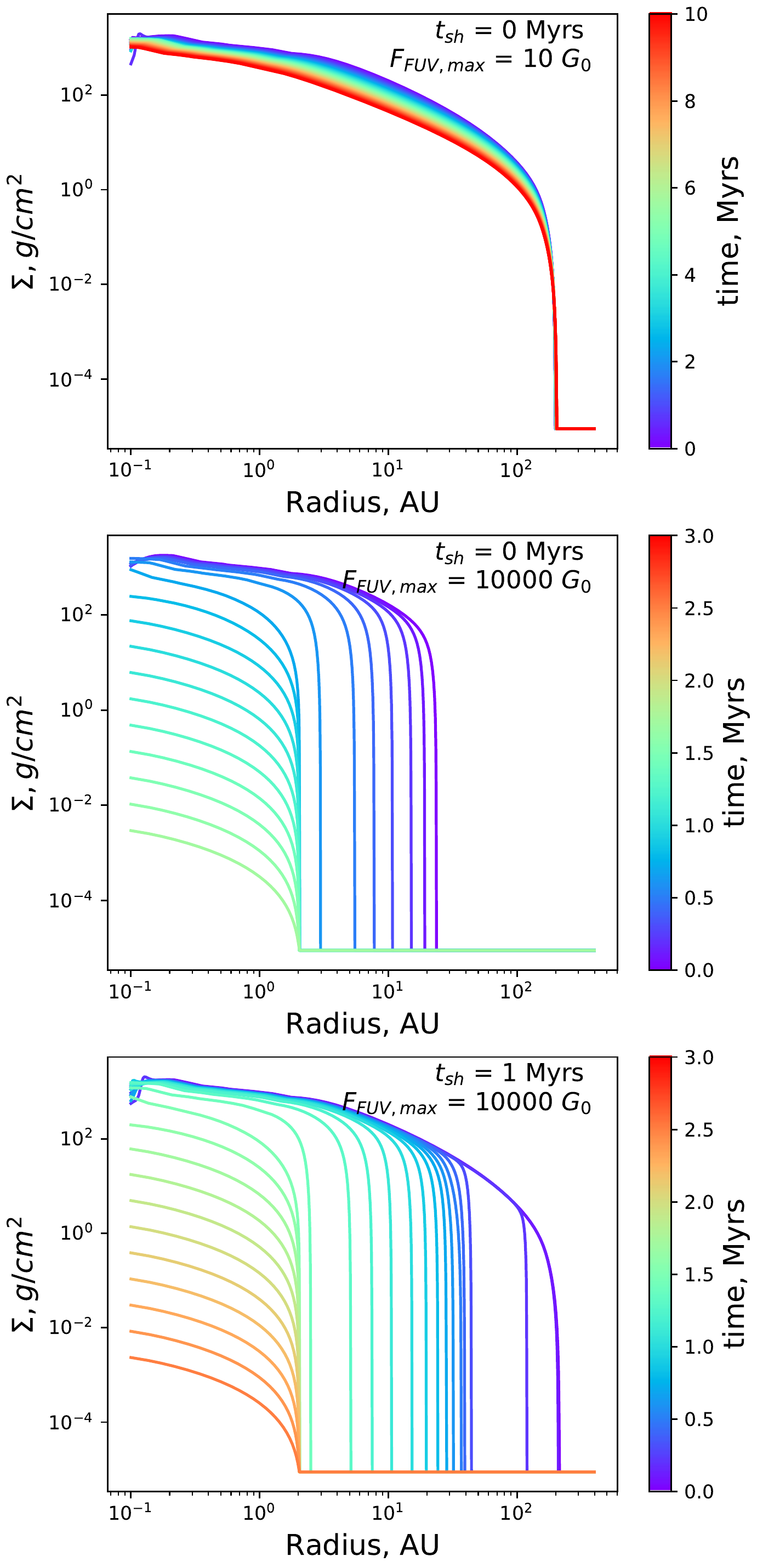}
    \caption{Examples of disc gas surface density evolution. Top panel: surface density evolution for a disc under very weak FUV field radiation ($F_{\textrm{FUV, max}}$ = 10 $G_0$. Middle panel: surface density evolution for a disc under very strong FUV field radiation  ($F_{\textrm{FUV, max}}$ = 10000 $G_0$) without shielding. Bottom panel: surface density evolution for a disc under very strong FUV field radiation  ($F_{\textrm{FUV, max}}$ = 10000 $G_0$) but with 1 Myrs of shielding.} 
    \label{fig:surface_density}
\end{figure}




\bsp	
\label{lastpage}
\end{document}